\documentclass[prd,a4paper, nofootinbib, preprintnumbers,amsmath, multicol]{revtex4}
\pdfoutput=1
\usepackage{latexsym}
\usepackage{epsfig}
\usepackage{amsmath}
\usepackage{amssymb}
\usepackage{graphicx}
\usepackage{bm}
\RequirePackage[colorlinks=true
,urlcolor=blue
,anchorcolor=blue
,citecolor=blue
,filecolor=blue
,linkcolor=blue
,menucolor=blue
,pagecolor=blue
,linktocpage=true
,pdfproducer=medialab
]{hyperref}
\makeatletter
\newcommand{\beq}{\begin{equation}}
\newcommand{\eeq}{\end{equation}}
\newcommand{\bea}{\begin{eqnarray}}
\newcommand{\eea}{\end{eqnarray}}
\newcommand{\ba}{\begin{array}}
\newcommand{\ea}{\end{array}}
\newcommand{\bi}{\begin{itemize}}
\newcommand{\ei}{\end{itemize}}
\newcommand{\bn}{\begin{enumerate}}
\newcommand{\en}{\end{enumerate}}
\newcommand{\bc}{\begin{center}}
\newcommand{\ec}{\end{center}}

\newcommand{\la}{\lambda}

\newcommand{\cO}{\mathcal{O}}

\begin{document}
\preprint{\footnotesize CERN-PH-TH/12-184}
\preprint{\footnotesize RM3-TH/12-12}
\preprint{\footnotesize DFPD-2012/TH/7}
\preprint{\footnotesize CP3-Origins-2012-018}
\preprint{\footnotesize DIAS-2012-19}
\preprint{\footnotesize TUM-HEP-843/12}

\title{\Large \color{red} \bf  Repressing Anarchy in Neutrino Mass Textures}

\vskip 1 cm

\author{Guido Altarelli $^{\color{blue}{a)}\color{blue}{b)}}$}
\email{guido.altarelli@cern.ch}
\author{Ferruccio Feruglio $^{\color{blue}{c)}}$}
\email{feruglio@pd.infn.it}
\author{Isabella Masina $^{\color{blue}{d)} \color{blue}{e)}}$}
\email{masina@fe.infn.it}
\author{Luca Merlo $^{\color{blue}{f)}\color{blue}{g)}}$}
\email{luca.merlo@ph.tum.de}

\affiliation{
$^{\color{blue}{a)}}$
{Dipartimento di Fisica `E.~Amaldi', Universit\`a di Roma Tre, \\INFN Sezione di Roma Tre, I-00146 Rome, Italy}\\
$^{\color{blue}{b)}}$
{CERN, Department of Physics, Theory Division, CH-1211 Geneva 23, Switzerland} \\
$^{\color{blue}{c)}}$
{Dipartimento di Fisica e Astronomia `G.~Galilei', Universit\`a di Padova, \\INFN Sezione di Padova, Via Marzolo~8, I-35131 Padua, Italy}\\
$^{\color{blue}{d)}}$ 
{Dipartimento di Fisica dell'Universit\`a di Ferrara, \\INFN Sezione di Ferrara, Via Saragat 1, I-44100 Ferrara, Italy}\\
$^{\color{blue}{e)}}$  
{CP$^\mathbf 3$-Origins and DIAS, SDU University, \\Campusvej 55, DK-5230 Odense M, Denmark}\\ 
$^{\color{blue}{f)}}$
{Physik-Department, Technische Universit\"at M\"unchen, \\James-Franck-Strasse, D-85748 Garching, Germany}\\
$^{\color{blue}{g)}}$
{TUM Institute for Advanced Study, Technische Universit\"at M\"unchen, \\Lichtenbergstrasse 2a, D-85748 Garching, Germany}
}

\baselineskip=15pt
\setcounter{page}{1}

\vskip 2cm
\begin{abstract}
The recent results that $\theta_{13}$ is relatively large, of the order of the previous upper bound, and the indications of a sizable deviation of $\theta_{23}$ from the maximal value are in agreement with the predictions of Anarchy in the lepton sector. The quark and charged lepton hierarchies can then be reproduced in a $SU(5)$ GUT context by attributing non-vanishing $U(1)_{FN}$ charges, different for each family, only to the $SU(5)$ tenplet states. The fact that the observed mass hierarchies are stronger for up quarks than for down quarks and charged leptons supports this idea. As discussed in the past, in the flexible context of $SU(5)\otimes U(1)_{FN}$, different patterns of charges can be adopted going from Anarchy to various types of hierarchy. We revisit this approach by also considering new models and we compare all versions to the present data. As a result we confirm that, by relaxing the ansatz of equal $U(1)_{FN}$ charges for all $SU(5)$ pentaplets and singlets, better agreement with the data than for Anarchy is obtained without increasing the model complexity. We also present the distributions obtained in the different models for the Dirac CP-violating phase. Finally we discuss the relative merits of these simple models.
\end{abstract}

\maketitle

%
%

\section{Introduction}

Recently our knowledge of the neutrino mixing matrix has much improved with the rather precise measurement of $\theta_{13}$ and the indication that $\theta_{23}$ is not maximal \cite{Fogli:2012ua,Tortola:2012te}(some constraints on the CP-violating phase in neutrino oscillations are also starting to emerge).
The rather large measured value of $\theta_{13}$ \cite{Abe:2011sj,Abe:2011fz,An:2012eh,Ahn:2012nd}, close to the old CHOOZ bound \cite{Apollonio:1999ae} and to the Cabibbo angle, and the indication that $\theta_{23}$ is not maximal both go in the direction of models based on Anarchy \cite{Hall:1999sn,Haba,deGouvea:2003xe}, i.e. of no symmetry in the leptonic sector, only chance (this possibility has been recently reiterated, for example, in Ref. \cite{deGouvea:2012ac}). Anarchy can be formulated in a $SU(5) \otimes U(1)_{FN}$ context by taking different Froggatt-Nielsen \cite{Froggatt:1978nt} charges only for the $SU(5)$ tenplets (for example $10\sim(a,b,0)$, where $a > b > 0$ is the charge of the first generation, b of the second, zero of the third) while no charge differences appear in the $\bar 5$ (e. g. $\bar 5\sim (0,0,0)$). If not explicitly stated, the Higgs fields are taken neutral under $U(1)_{FN}$. The $SU(5)$ generators act
ÔverticallyÕ inside one generation, whereas the $U(1)_{FN}$ charges are different ÔhorizontallyÕ from one
generation to the other. If, for a given interaction vertex, the $U(1)_{FN}$  charges do not add to zero, the
vertex is forbidden in the symmetric limit. However, the $U(1)_{FN}$ symmetry (that we can assume to be a gauge symmetry\footnote{Gauge anomalies can be cancelled adding a set of heavy additional fermions, vector-like under $SU(5)$ and chiral under $U(1)_{FN}$. They have no impact on the present discussion.}) is spontaneously broken by
the VEVs $v_f$ of a number of ÔflavonÕ fields with non-vanishing charges and GUT-scale masses. Then a forbidden coupling
is rescued, but is suppressed by powers of the small parameters $\lambda = v_f/M$, with $M$ a large mass, with the exponents larger for larger charge mismatch. Thus the charges fix the powers of $\lambda$, hence the degree of suppression of all elements of mass matrices, while arbitrary coefficients $c_{ij}$ of order 1 in each entry of mass matrices are left unspecified (so that the number of parameters exceeds the number of observable quantities). A random selection of these $c_{ij}$ parameters leads to distributions of resulting values for the measurable quantities. For Anarchy ($A$) the mass matrices in the leptonic sector are totally random; on the contrary, in the presence of non-vanishing charges different entries carry different powers of the order parameter and thus some hierarchies are enforced. There are many variants of these models: fermion charges can all be non-negative with only negatively charged flavons, or there can be fermion charges of different signs with either flavons of both charges or only flavons of one charge. In models with no SeeSaw, the $\bar 5$ charges completely fix the hierarchies (or Anarchy, if the case) in the neutrino mass matrix.  If Right-Handed (RH) neutrinos are added, they transform as $SU(5)$ singlets and can in principle carry $U(1)_{FN}$ charges, which also must be all equal in the Anarchy case. With RH neutrinos the SeeSaw mechanism can take place and the resulting phenomenology is modified. 

Anarchy and its variants, all sharing the dominance of ramdomness in the lepton sector, are to be confronted with models based on discrete flavour groups. These more ambitious models are motivated by the fact that the data suggest some special mixing patterns as good first approximations like Tri-Bimaximal (TB) or Golden Ratio (GR) or Bi-Maximal (BM) mixing, for example. The corresponding mixing matrices all have $\sin^2{\theta_{23}}=1/2$, $\sin^2{\theta_{13}}=0$, values that are good approximations to the data (although less so since the most recent data), and differ by the value of the solar angle $\sin^2{\theta_{12}}$. The observed $\sin^2{\theta_{12}}$, the best measured mixing angle,  is very close, from below, to the so called Tri-Bimaximal (TB) value \cite{Harrison:2002er,Harrison:2002kp,Xing:2002sw,Harrison:2002et,Harrison:2003aw} of $\sin^2{\theta_{12}}=1/3$. Alternatively, it is also very close, from above, to the Golden Ratio (GR) value \cite{Kajiyama:2007gx,Everett:2008et,Ding:2011cm,Feruglio:2011qq} $\sin^2\theta_{12}=1/\sqrt{5}\,\phi \sim 0.276$, where $\phi= (1+\sqrt{5})/2$ is the GR (for a different connection to the GR, see Refs.~\cite{Rodejohann:2008ir,Adulpravitchai:2009bg}). On a different perspective, one has also considered models with Bi-Maximal (BM) mixing \cite{Vissani:1997pa,Barger:1998ta,Nomura:1998gm,Altarelli:1998sr}, where $\sin^2{\theta_{12}}=1/2$, i.e. also maximal, as the neutrino mixing matrix before diagonalization of charged leptons.  One can think of models where a suitable symmetry enforces BM mixing in the neutrino sector at leading order (LO) and the necessary, rather large, corrective terms to $\theta_{12}$ arise from the diagonalization of the charged lepton mass matrices \cite{Raidal:2004iw,Minakata:2004xt,Minakata:2005rf,Frampton:2004vw,Ferrandis:2004vp,Kang:2005as,Altarelli:2004jb,Li:2005ir,Cheung:2005gq,Xing:2005ur,Datta:2005ci,Ohlsson:2005js,Antusch:2005ca,Lindner:2005pk,King:2005bj,Masina:2005hf,Dighe:2006zk,Chauhan:2006im,Schmidt:2006rb,Hochmuth:2006xn,Plentinger:2006nb,Plentinger:2007px,Altarelli:2009gn,Toorop:2010yh,Patel:2010hr,Meloni:2011fx,Shimizu:2010pg,Ahn:2011yj}. Thus, if one or the other of these coincidences is taken seriously, models where TB or GR or BM mixing is naturally predicted provide a good first approximation (but these hints cannot all be relevant and it is well possible that none is).
The corresponding mixing matrices have the form of rotations with fixed special angles. Thus one is naturally led to discrete flavour groups. Models based on discrete flavour symmetries, like $A_4$ or $S_4$, have been proposed in this context and widely studied  \cite{Altarelli:2010gt,Ishimori:2010au,Grimus:2010ak,Parattu:2010cy,Grimus:2011fk,Altarelli:2012ss,Bazzocchi:2012st}. In these models the starting Leading Order (LO) approximation is completely fixed (no chance), but the Next to LO (NLO) corrections still introduce a number of undetermined parameters, although in general much less numerous than for $U(1)_{FN}$ models. These models are therefore more predictive and typically, in each model, one obtains relations among the departures of the three mixing angles from the LO patterns, restrictions on the CP violation phase $\delta$, mass sum rules among the neutrino mass eigenvalues, definite ranges for the neutrinoless-double-beta decay effective Majorana mass and so on. 

The aim of this note on $U(1)_{FN}$ models is to update, on the basis of the present more precise data, our old analysis \cite{Altarelli:2002sg} that shows that, even if one accepts a mainly chaotic approach to lepton mixing, the Anarchy ansatz is perhaps oversimplified and that suitable differences of $U(1)_{FN}$ charges, which must in any case be present for tenplets, if also introduced within pentaplets and singlets, actually lead to distributions that are in much better agreement with the data with the same number of random parameters. In fact Anarchy can be improved by implementing mechanisms that enforce the relative smallness of $\theta_{13}$ and of $r=\Delta m^2_{solar}/\Delta m^2_{atm}$.  The first goal can be achieved by restricting Anarchy only to the $\mu-\tau$ (or 2-3) sector, by taking the pentaplet charges as $\bar 5\sim (c,0,0)$, $c > 0$, (``$\mu-\tau$ Anarchy'', $A_{\mu\tau}$) \cite{Buchmuller:1998zf,Buchmuller:2011tm}.  Both improvements can be realised by taking $\bar 5\sim (c,d,0) $, $c >d > 0$, (this ``hierarchical'', H, pattern was not considered in Ref. \cite{Altarelli:2002sg}). In each case the tenplet charges can be readjusted in order to maintain the correct ratios of charged fermion masses. We only consider models with normal hierarchy (NH) here, because, as shown in Ref. \cite{Altarelli:2002sg}, in this framework models with inverse hierarchy (IH) tend to favour a solar angle close to maximal.
In models with no SeeSaw, the $\bar 5$ charges completely fix the hierarchies (or the Anarchy) in the neutrino mass matrix, through the dimension-5 Weinberg operator, $m_\nu = \Psi_{\bar 5}^T \Psi_{\bar 5}\,H_u\,H_u/M$. The distributions arising from the models $A$, $A_{\mu\tau}$ and $H$ with no SeeSaw can directly be obtained and compared with the data and we shall see to which extent $H$ is better than $A_{\mu\tau}$ which, in turn, is better than $A$.
If RH neutrinos are added, they transform as $SU(5)$ singlets and can in principle carry $U(1)_{FN}$ charges, which also are all equal in the Anarchy case. With RH neutrinos the SeeSaw mechanism can take place and the resulting phenomenology is modified. It is easy to show that models with all non-negative charges and one single flavon have particularly simple factorization properties \cite{Altarelli:2004za}. In particular, in the SeeSaw expression for the light neutrino mass matrix $m_\nu = m_D^T\,M^{-1}\,m_D$, with $m_D$ and $M$ denoting the neutrino Dirac and Majorana mass matrices, respectively, the dependence on the RH charges drops out in this case and only that from the $\bar 5$ remains. In these simplest models the only difference between the version with and without SeeSaw, for each model $A$, $A_{\mu\tau}$ and $H$ is that the extraction procedure for the random numbers is different: in the no SeeSaw version the entries of the neutrino mass matrix $m_\nu$ are directly generated while in the SeeSaw case the extraction is done for $m_D$ and $M$ and then $m_\nu$ is derived by the SeeSaw formula. For example, in Anarchy models the smallness of $r$ is to some extent reproduced by the spreading of the mass distribution resulting from the product of three mass matrices. Models with naturally large 23 neutrino mass splittings (so that $r$ is small) are obtained if we allow negative charges and, at the same time, either introduce flavons of opposite charges or allow that matrix elements with overall negative charge are vanishing. For example, one can take $\bar 5\sim (c,0,0)$ like in ``$\mu-\tau$'' Anarchy and $1 \sim (e, -e, 0)$, $e > 0$, with two flavons of opposite charges, with equal VEV, and SeeSaw (we denote this model as Pseudo $\mu \tau$-Anarchy,``$PA_{\mu \tau}$''). The ``lopsided'' structure of $\bar 5\sim (c,0,0)$ results in naturally small 23 subdeterminant in the neutrino mass matrix after SeeSaw and to $r$ naturally small.

In the following we discuss in more detail the models, the procedure of extraction of the random coefficients and the value of the expansion parameter $\lambda$ that maximizes the success rate for each model. We then discuss the mass distributions and mixing angles that we obtain and, finally, we compare these distributions with the data. The conclusion is that the most effective model is $H$ in the no SeeSaw case and $PA_{\mu \tau}$ in the SeeSaw case, while  $A_{\mu\tau}$ and $A$ follow in the order and are much less successful.

%
%

\section{Models and Results}

In the following analysis we adopt the results of the fit of Ref.~\cite{Fogli:2012ua} (see also Ref.~\cite{Tortola:2012te}). 
The $2 (3) \sigma$ ranges in the case of NH are:
\beq
\begin{gathered}
7.15 (6.99) \times 10^{-5} {\rm eV}^2 \le \Delta m^2_{solar} \le  8.00(8.18) \times 10^{-5} {\rm eV}^2 \\ 
2.27(2.19) \times 10^{-3} {\rm eV}^2 \le \Delta m^2_{atm} \le  2.55(2.62) \times 10^{-3} {\rm eV}^2 \\ 
0.0193(0.0169)<\sin^2\theta_{13} < 0.0290(0.0313) \\ 
0.275(0.259) <\sin^2\theta_{12}<0.342(0.359)\\
0.348(0.331) <\sin^2\theta_{23}<0.448(0.637)
\end{gathered}
\label{F12}
\eeq
where $\Delta m^2_{solar} =m_2^2-m_1^2$ and $\Delta m^2_{atm} =m_3^2-(m_2^2+m_1^2)/2$. 

We consider models with different patterns - Anarchy, genuine or Pseudo $\mu\tau$-Anarchy, and hierarchy - induced by a $U(1)$ flavour symmetry \cite{Froggatt:1978nt}: we present the transformation properties of all the fields in table \ref{tab-models}, in a notation that corresponds to the $SU(5)$ GUT embedding. In the non-SeeSaw case, the neutrino mass matrix $m_\nu$ is generated via the effective Weinberg operator $ \Psi_{\bar 5}^T \Psi_{\bar 5}\,H_u\,H_u/M$. In the SeeSaw case the flavour charges determine the Dirac and RH Majorana mass matrices, $m_D$ and $M$, which give the effective neutrino mass $m_\nu=m_D^T M^{-1} m_D$ at low energy. If the RH neutrino charges all have the same sign and there is a single flavon, it is known that the structure of $m_\nu$ in powers of $\lambda$ is the same as for the non-SeeSaw case \cite{Irges:1998ax}. The coefficients in front of $\lambda$ are randomly generated complex numbers $c=|c| e^{i \phi_c}$. In the spirit of the $U(1)$ flavour symmetry, $|c| = {\cal O}(1)$ while the phase $\phi_c$ is arbitrary. 

\begin{table}[!t]
\vspace{0.4cm}
\begin{center}
\begin{tabular}{|c|c|c|c|}
\hline 
& & & \\ 
{Model}& ${{\Psi_{10}}}$ & ${\Psi_{\bar 5}}$ & ${{\Psi_1}}$ \\ 
& & & \\
\hline
\hline
& & & \\ 
{Anarchy ($A$)}& (3,2,0)& (0,0,0) & (0,0,0)\\ 
& & & \\
\hline
 & & & \\ 
{$\mu\tau$-Anarchy ($A_{\mu\tau} $)}& (3,2,0) & (1,0,0) & (2,1,0)\\ 
& & & \\
\hline
& & &  \\ 
{Pseudo $\mu\tau$-Anarchy  ($PA_{\mu \tau} $)}& (5,3,0) & (2,0,0) & (1,-1,0) \\ 
& & &  \\ 
\hline
& & &  \\ 
{Hierarchy ($H$)}& (5,3,0) & (2,1,0) & (2,1,0) \\ 
& & &  \\
\hline
\end{tabular}
\end{center}
\caption{Models and their flavour charges suitable for an implementation in a Supersymmetric (SUSY) $SU(5)\otimes U(1)_{\rm FN}$ GUT. The flavon charge is  $-1$. The charges of the $\Psi_{10}$ have been chosen to reproduce the mass hierarchies of the charged fermions, for the values of $\lambda$ that maximize the success rates for each model (see text). The Higgs $H_{u,d}$ charges have all been taken as vanishing in these models.}
\label{tab-models}
\end{table}

We report in Fig.~\ref{fig-SR} the per cent success probability for each model as a function of the expansion parameter $\lambda$. We defined $P= n_{ok}/n_{tot}$, where $n_{tot}$ is the total number of randomly generated models (typically much larger than $10^6$) and $n_{ok}$ is the number of models consistent with the $2\sigma$ ranges of Eq.~(\ref{F12}) \cite{Fogli:2012ua}. The relative error on $P$ is estimated to be $\cO(1/\sqrt{n_{ok})} $. The thickness of the lines indicates this statistical error. Since the success rate depends on the window selected to extract the random coefficients, to estimate the ambiguity from this effect we flatly generated $|c|$ in the interval $[0.5,2]$ (solid shaded) and also in the interval $[0.8,1.2]$ (dashed shaded). The phases $\phi_c$ have been chosen to be flatly distributed in $[0,2 \pi]$. We also checked the stability of the results adopting a gaussian distribution for $|c|$: in particular, the results for $|c|$ in the interval $[0.5,2]$ ($[0.8, 1.2]$) are exactly reproduced by a gaussian distribution with central value $1.2$ ($1$) and standard deviation $0.4$ ($0.1$). Furthermore, we have also considered flat distributions for real and maginary parts of $c$, letting $c$ vary in a square centered at the origin of the complex plane. We have studied the dependence on the size of the square. We got slightly different results for the success probabilities, for the value of the parameter $\lambda$ that maximizes the success rate and for the distributions of the various observables. However, the relative ability of the different models to fit the data, that we consider the most important outcome of the present analysis, is stable and independent from the distributions we used to generate the input coefficients $c$.

The flavour charges of $\Psi_{10}$ in table \ref{tab-models} have been chosen in order to reproduce the mass hierarchies of the charged fermions for the value of $\lambda$ that for each model maximizes the success rates. We scanned for $\Psi_{10}$ integer charges $(a,b,0)$ with $0<b \le a \le 10$ and $b\le 5$. For each choice of charges we compared the distributions of the six mass ratios $m_e/m_\mu$, $m_\mu/m_\tau$, $m_d/m_s$, $m_s/m_b$, $m_u/m_c$ and $m_c/m_t$ and of the three CKM matrix elements $V_{us}$, $V_{ub}$ and $V_{cb}$ with the corresponding 3$\sigma$ experimental interval, renormalized at the GUT scale assuming $\tan\beta=10$ (see for instance Ref.~\cite{Altarelli:2010at}). 
We extracted the most successful charges by asking that the experimental interval of each parameter overlaps with the 1$\sigma$ region of the theoretical distribution. Only for the $PA_{\mu \tau}$ model, this requirement is fulfilled for all the nine observables. For all the other models, there is a tension between the choices which fit $m_e/m_\mu$, which favour large values of $a-b$, and the choices that reproduce $V_{us}$, which require $a-b$ small. In these models, a tuning of the unknown order one parameters is needed in order to reproduce both these quantities. We gave our preference to the solutions which correctly reproduce $V_{us}$. For a given model, the solution we found is not unique and other choices of the charges are equally successful, provided we change the size of the expansion parameter $\lambda$: indeed, redefining $\lambda\to \lambda^{1/q}$, $a\to q\,a$ and $b\to q\,b$, where $q$ is a positive number, the structure of the mass matrices that we will show in the following does not change. This observation alleviates the otherwise arbitrary choice of integer charges. In table I we list a representative set of charges that passed our test. Choosing a vanishing charge for both the $10$ and the $\bar 5$ SU(5) representations of the third generation have no impact on our selection procedure, that relies on quantities that are only sensitive to charge differences. The nominal values of the charges listed in table I require a large value of $\tan\beta$, since the top and bottom Yukawa couplings are both expected to be of order one. Smaller values of $\tan\beta$ can be easily accommodated by allowing for a positive charge of $H_d$.

\begin{figure}[t!]
\begin{center}
\begin{tabular}{c}
\includegraphics[width=6.9cm]{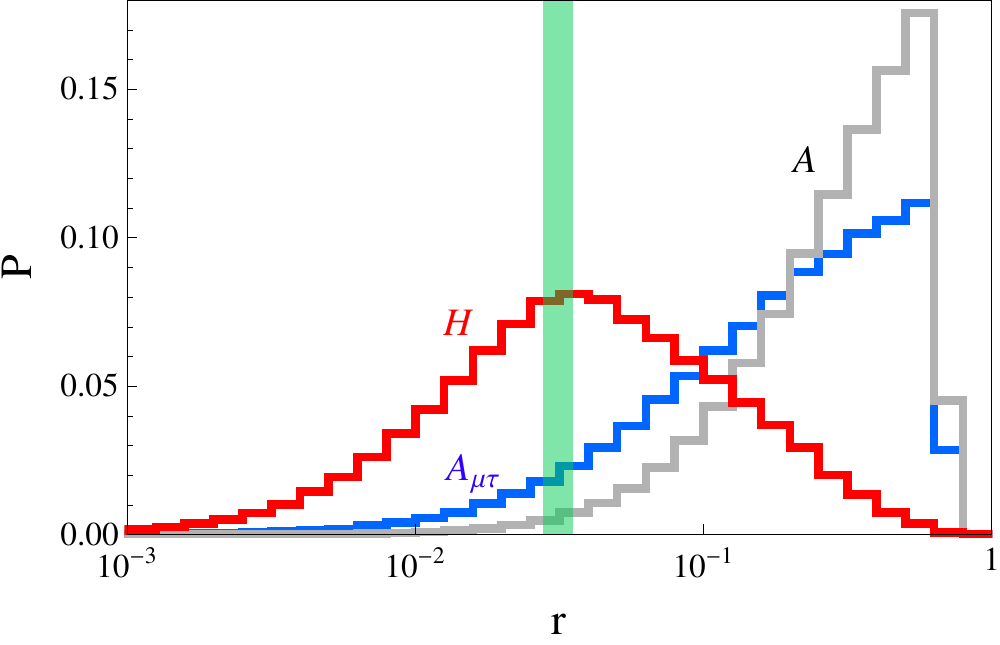}     \,\includegraphics[width=6.9cm]{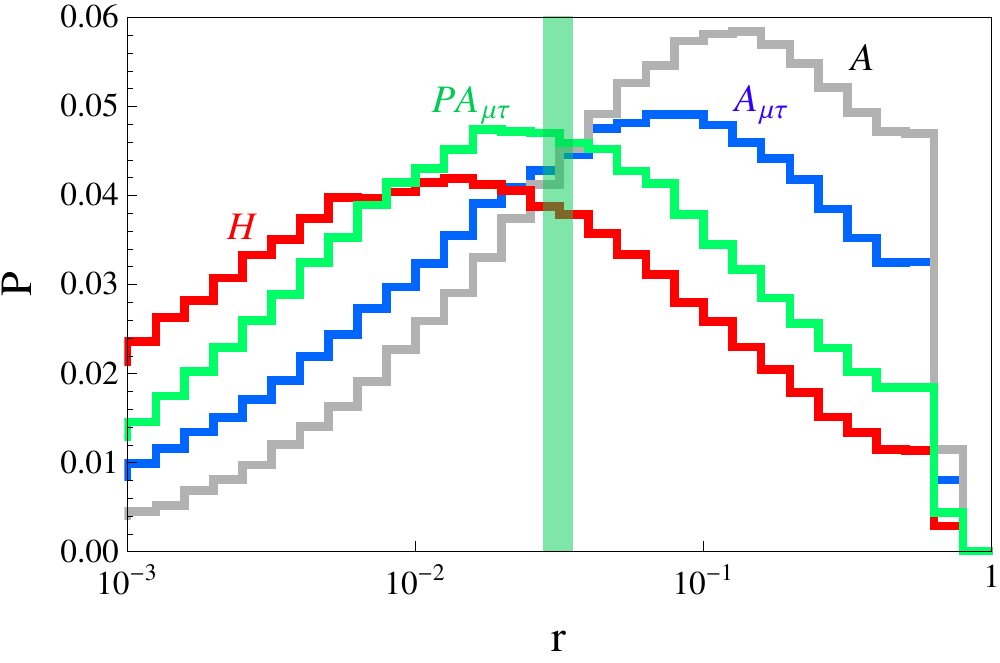}  \\\,\\
\includegraphics[width=6.9cm]{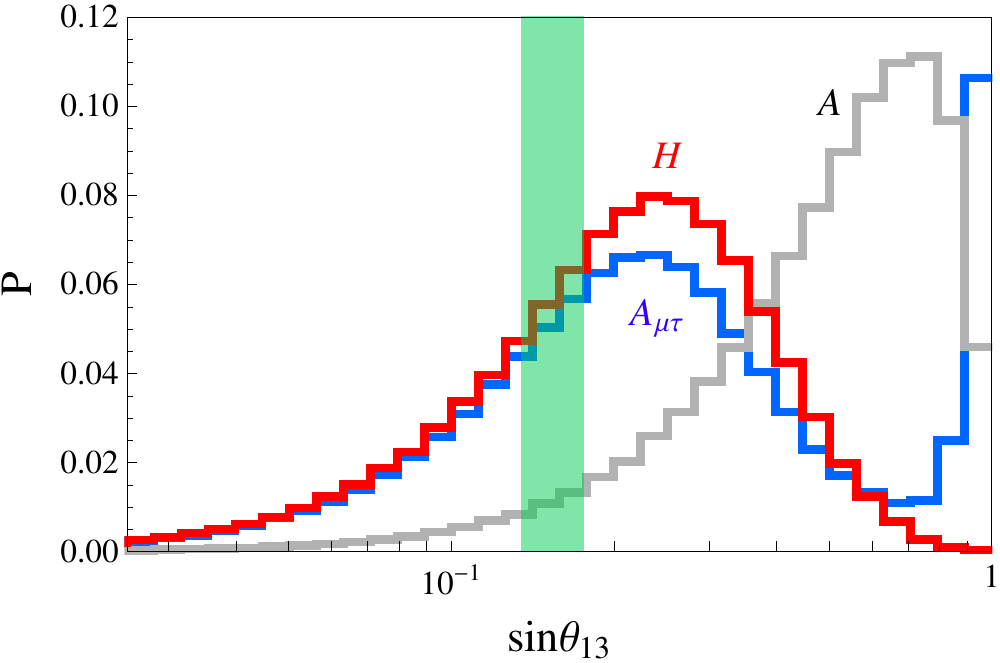}  \, \includegraphics[width=6.9cm]{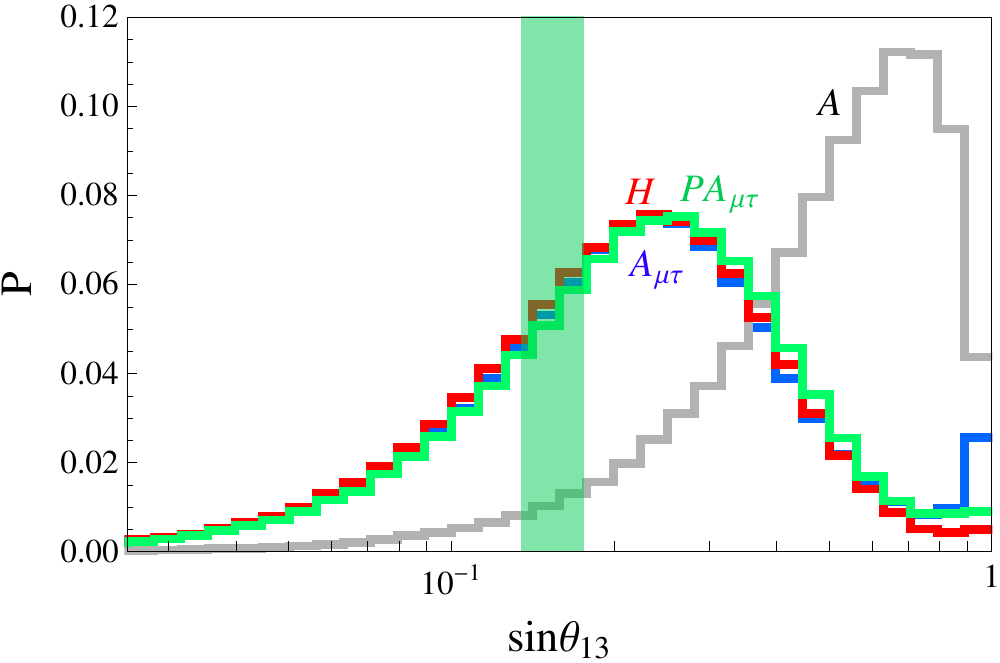} \\\,\\
\includegraphics[width=6.9cm]{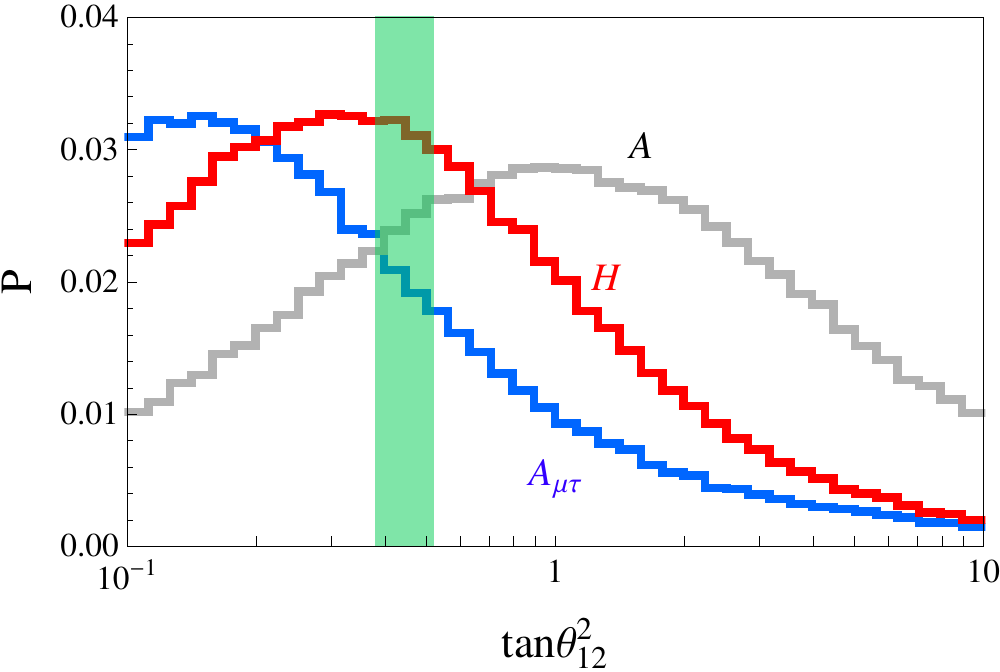}  \, \includegraphics[width=6.9cm]{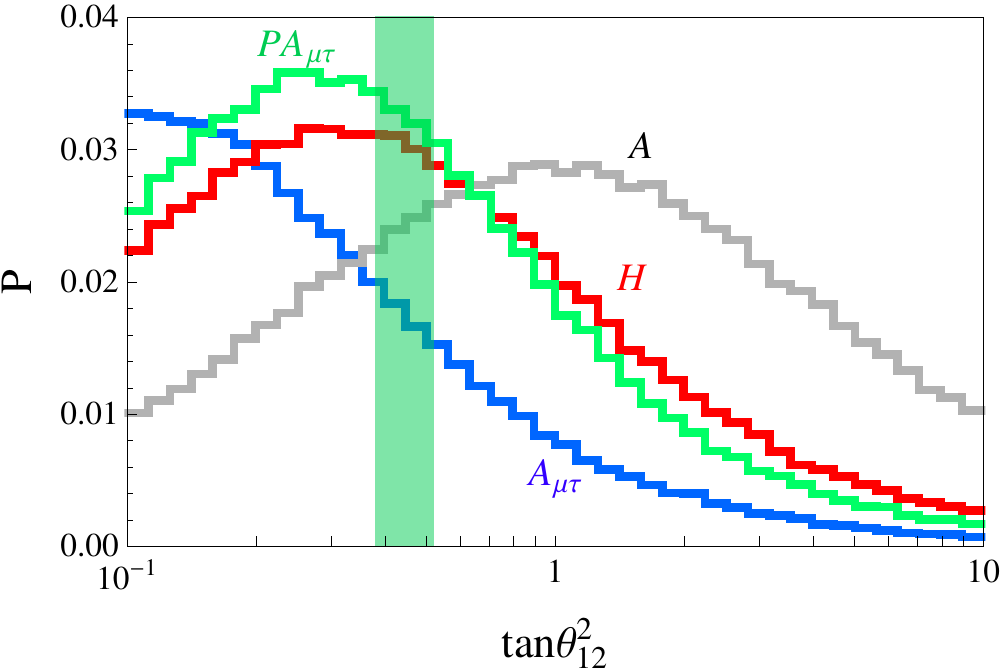} \\\,\\
\includegraphics[width=6.9cm]{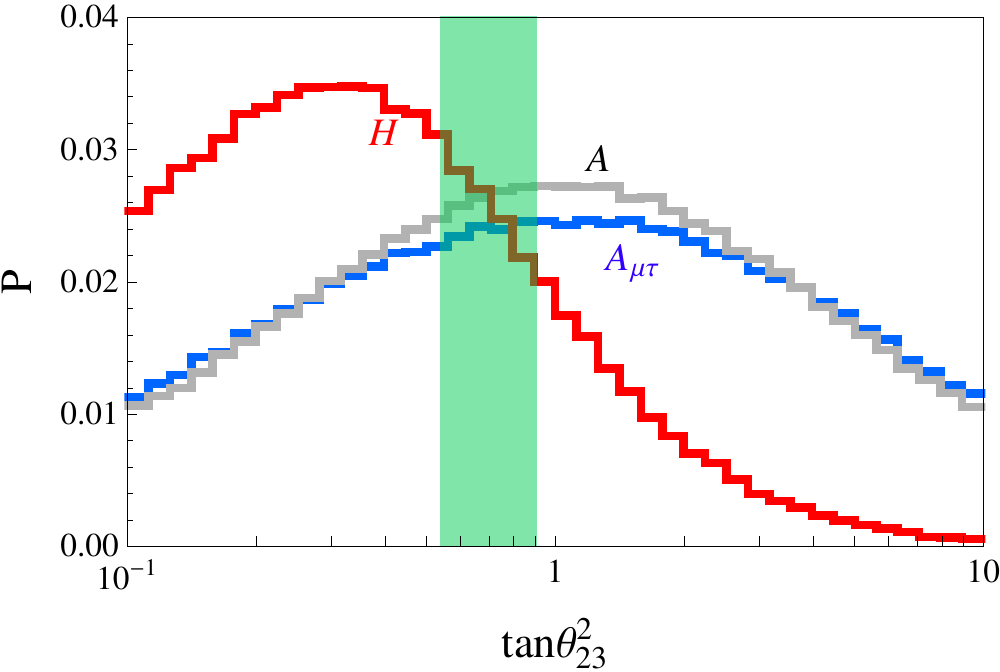}  \, \includegraphics[width=6.9cm]{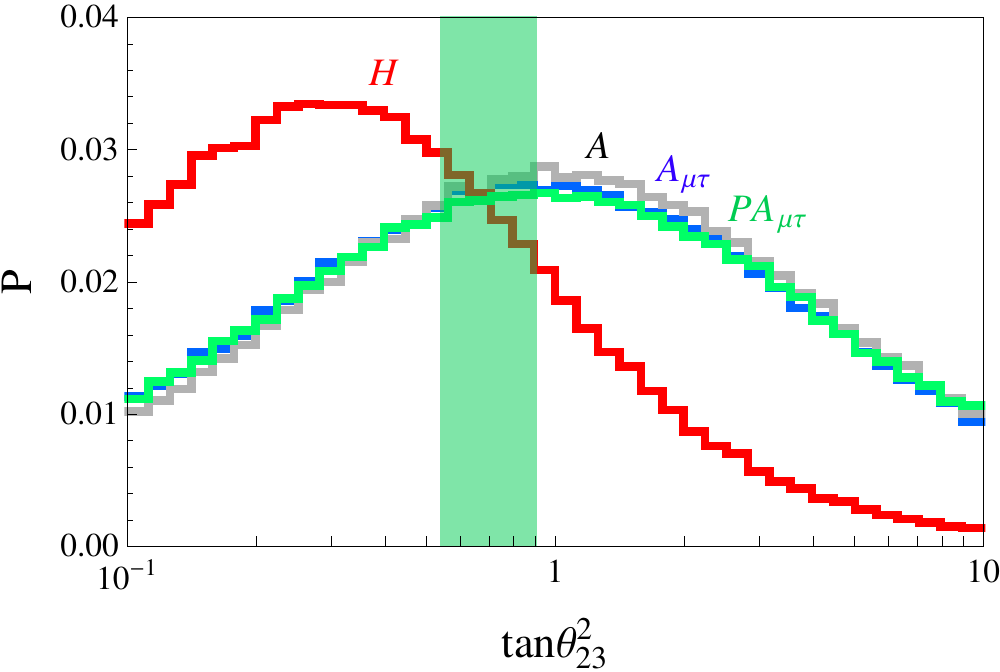}  
\end{tabular}
\end{center}\vspace*{-0.4cm} 
\caption{Probability distributions of $r, \sin\theta_{13}, \tan^2\theta_{12}, \tan^2 \theta_{23}$ without (left column) and with (right column) SeeSaw. 
The modulus (argument) of the complex random coefficients has been generated in the interval $[0.5,2]$ ($[0,2\pi]$) with a flat distribution.
For $A$ and $A_{\mu\tau}$ we considered $\lambda=0.2$, for $H$ and $PA_{\mu \tau}$ we considered $\lambda=0.4$. The shaded vertical band
emphasizes the experimental $2\sigma$ window according to \cite{Fogli:2012ua}.}
\label{fig-D}
\end{figure}

In Fig.~\ref{fig-D} we display the probability distributions for 
$r$, $\sin\theta_{13}$, $\tan^2\theta_{12}$ and $\tan^2\theta_{23}$, fixing $\lambda$ at a representative value and selecting $|c|\in [0.5,2]$. These distributions have been normalized so that the integrated probability 
is equal to unity. Notice that we use a logarithmic scale. The (green) shaded vertical regions refer to the experimental data at $2\sigma$ \cite{Fogli:2012ua}.

We now discuss each model in turn. 

\begin{itemize}
\item{Anarchy} ($A$) \cite{Hall:1999sn,deGouvea:2003xe,deGouvea:2012ac}. Neglecting the randomly generated coefficients, the texture of the mass matrices for charged leptons and neutrinos (with or without SeeSaw), expressed in powers of $\lambda$, reads explicitly\beq
m_{\ell}= \left(  \begin{matrix}  \la^3 &    \la^3 &   \la^3 \\  \la^2 &    \la^2 &   \la^2 \\ 1 &   1 &   1 \end{matrix}     \right)\,,\qquad\qquad
m_{\nu}= \left(  \begin{matrix}  1 &    1 &   1 \\  1 &   1 &  1 \\ 1 &   1 &   1 \end{matrix}     \right)\,.
\label{A}
\eeq
The success rate for neutrino masses and mixing angles is independent of $\la$. We then choose $\la = 0.2 - 0.3$, which ensure a reasonable hierarchy for charged fermions according to the charges selected for the $\Psi_{10}$ representation. As can be seen from Fig.~\ref{fig-SR}, the success rate is quite small in both the no SeeSaw and SeeSaw cases: for the no SeeSaw case the success rate has been multiplied by a factor of $10$ to facilitate its comparison with the other models. The reasons for such a modest performance can be understood by inspecting Fig.~\ref{fig-D}: the most severe problem is the prediction of a too large $\theta_{13}$ and, in the no SeeSaw case, also of a too large value of $r$. In the SeeSaw case the latter problem is cured by the spreading of neutrino mass eigenvalues produced by the product of three random matrix factors. As for the mixing angles $\theta_{ij}$ the distributions are all similar and, with a logarithmic scale, appear peaked near $\pi/4$.

\begin{figure}[t!]
\begin{center}
\includegraphics[width=7.3cm]{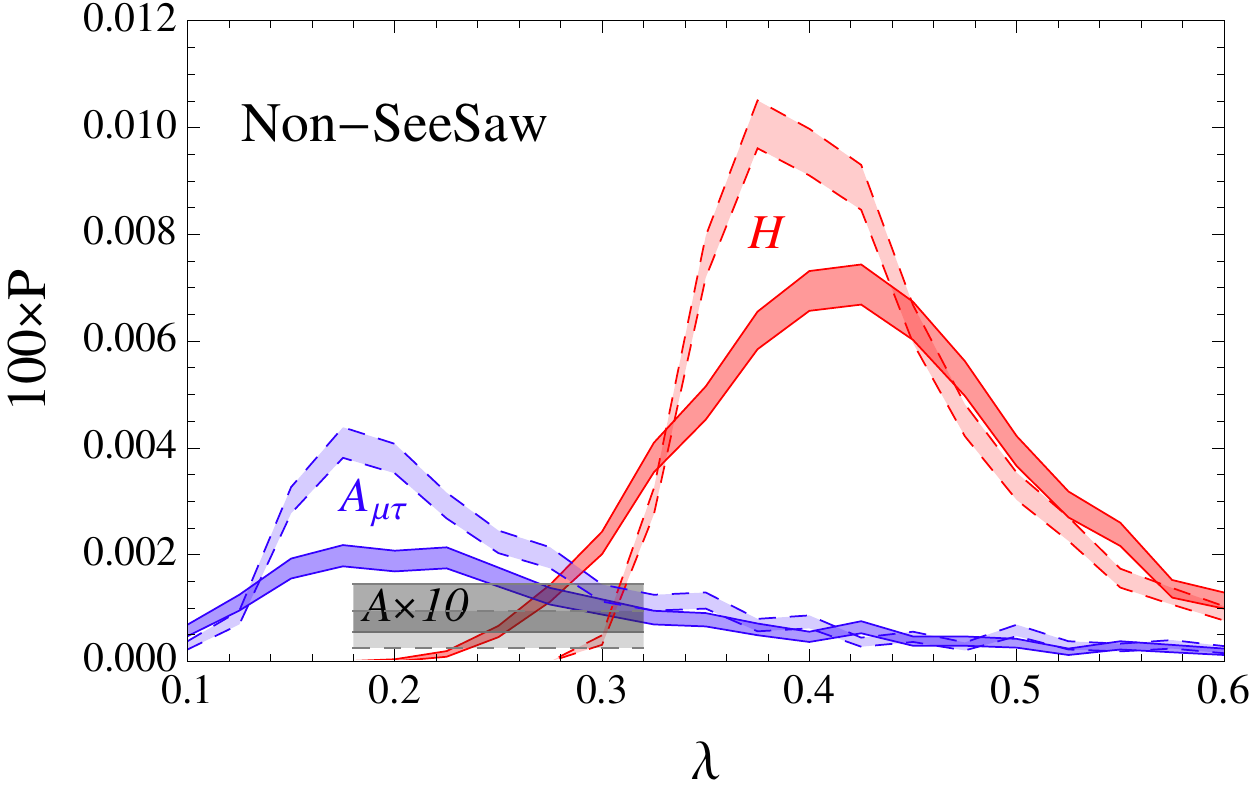}  \qquad
\includegraphics[width=7.3 cm]{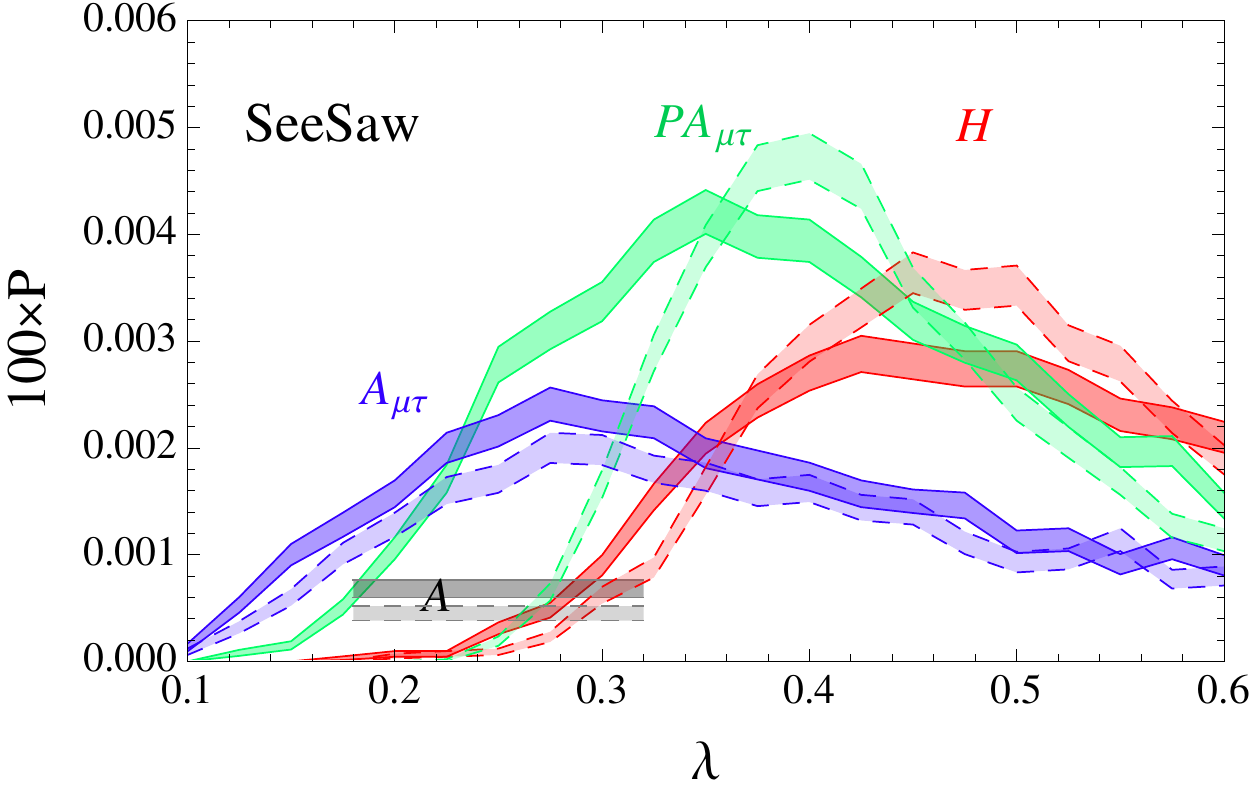} 
\end{center}
\caption{Per cent probability of success to fulfill the $2\sigma$ ranges of Eq.~(\ref{F12}) \cite{Fogli:2012ua} as a function of 
$\lambda$,  
without (left) and with SeeSaw (right). 
Solid (dashed) lines are obtained extracting $|c|$ with
a flat distribution in the interval $[0.5,2]$  ($[0.8,1.2]$). Phases $\phi_c$ are flatly distributed in $[0,2 \pi]$.
The thickness of the curves represent the statistical error, estimated as discussed in the text.
In the left plot, the success rate for $A$ has been multiplied by a factor of $10$.}
\vspace*{0.5cm} 
\label{fig-SR}
\end{figure}

\item{$\mu\tau$-Anarchy ($A_{\mu\tau}$) \cite{Buchmuller:1998zf,Masina:2005am,Buchmuller:2011tm}}. In this case only the $\mu\tau$ block of $m_\nu$ is anarchical
\beq
m_{\ell}= \left(  \begin{matrix}  \la^4 &    \la^3 &   \la^3 \\  \la^3 &    \la^2 &   \la^2 \\ \la &   1 &   1 \end{matrix}     \right)\,,\qquad\qquad
m_{\nu}= \left(  \begin{matrix}  \la^2 &    \la &   \la \\  \la &   1 &  1 \\ \la &   1 &   1 \end{matrix} \right)  
\label{SA-23}
\eeq
and the success rate is maximized for $\la\sim0.2$ and $\la\sim0.28$ for the no SeeSaw and SeeSaw cases respectively. 
In both cases the performance of $A_{\mu\tau}$ is better than $A$. 
The main problem of the $A_{\mu\tau}$ model is the
prediction of a too small $\theta_{12}$ and, in the no SeeSaw case, also a too large value for $r$. If by accident the 22 matrix element of $m_\nu$ is numerically of order $\la$ then $\theta_{12} \sim \cO(1)$ and $\sqrt{r}\sim \cO(\la)$: with a single fine tuning one fixes both problems.

\item{Pseudo $\mu\tau$-Anarchy ($PA_{\mu \tau}$) \cite{Altarelli:2002sg}}. This is a SeeSaw model with two flavons of opposite charges and equal VEVs. The 2 and 3 entries of the pentaplets have the same charges, but the 1 and 2 RH neutrinos have opposite charges. The result is that $m_\nu$ displays an apparently anarchical 23 sector,
\beq
m_{\ell}= \left(  \begin{matrix}  \la^7 &    \la^5 &   \la^5 \\  \la^5 &    \la^3 &   \la^3 \\ \la^2 &   1 &   1 \end{matrix}     \right) \,, \,\,\,\, 
m_D = \left(  \begin{matrix}   \la^3 & \la  & \la   \\  \la & \la  & \la  \\ \la^2  & 1   & 1   \end{matrix}     \right) \, , \,\,\,\,
M= \left(  \begin{matrix}   \la^2 &  1 & \la   \\ 1 & {\la}^2   & \la  \\ \la &  \la  &  1  \end{matrix}     \right)  \, , \,\,\,\,
m_{\nu}= \left(  \begin{matrix}  \la^4 & \la^2  & \la^2   \\ \la^2 & 1   &  1 \\  \la^2 & 1   &  1  \end{matrix}     \right)  \, ,
\eeq
but the associated coefficients automatically induce a suppression of the $\mu\tau$ determinant\footnote{Without the RH neutrinos, $PA_{\mu \tau}$ model corresponds to the $A_{\mu\tau}$, as indeed this suppression mechanism for the determinant does not hold.}, which is desirable to justify 
the smallness of $r$ while $\theta_{12} \sim \cO(1)$. The success rate for this model is maximized for $\la \sim0.35-0.4$.
For such values the distributions of $r$ and $\tan^2\theta_{12}$ in Fig.~\ref{fig-D} are indeed nearly centered in the experimental range.  
The mixing angle $\theta_{23}$ is instead naturally maximal and its distribution is indistinguishable with respect to $A$ and $A_{\mu\tau}$ models.
Notice that the $PA_{\mu\tau}$ model emerged as favorite in the 2005 update of the analysis of Ref.~\cite{Altarelli:2002sg}.

\item{Hierarchy ($H$).} Both for the SeeSaw and no SeeSaw cases, the charged lepton and neutrino mass matrices read
\beq
m_{\ell}= \left(  \begin{matrix}  \la^7 &    \la^6 &   \la^5 \\  \la^5 &    \la^4 &   \la^3 \\ \la^2 &   \la &   1 \end{matrix}     \right) \,,\qquad\qquad
m_{\nu}= \left(  \begin{matrix}  \la^4 &    \la^3 &   \la^2 \\  \la^3 &   \la^2 &  \la \\ \la^2 &   \la &   1 \end{matrix}     \right)\,.
\eeq

The success rate is maximized for $\la\sim0.4$ and $\la\sim0.45$ for the non-SeeSaw and SeeSaw cases, respectively.
For non-SeeSaw one finds a successful model every $10,000$ trials, that is a factor of 100 better than the Anarchy texture. For the SeeSaw the success rate is slightly lower, but still a factor of 10 better than Anarchy. This is mainly due to the hierarchical structure in both $12$ and $23$ sectors, that ensures a small $r \sim \lambda^4$ and   
$\tan^2 \theta_{12} \sim \tan^2 \theta_{23}\sim \sin\theta_{13} \sim \lambda^2$.
As can be seen from Fig.~\ref{fig-D}, with $\lambda = 0.4$, the maxima of the distributions of these observables are nicely close 
to their experimentally allowed range at $2\sigma$.
Notice that the distributions of $\theta_{12}$ and $\theta_{23}$ are similar. 
In particular, $\theta_{23}$ is peaked at a slightly smaller value than its present experimental $2\sigma$ range. 
Despite this, the rate of success of $\theta_{23}$ equals those of the three previously studied versions of anarchical models, namely
$A$, $A_{\mu\tau}$ and $PA_{\mu\tau}$.
\end{itemize}

We can now directly compare the four models. As from Fig.~\ref{fig-SR}, $H$ is the best performing model for the non-SeeSaw case, for values of 
$\lambda$ larger than about $0.3$, while for smaller values, $A_{\mu\tau}$ has the best success rate. This suggests that a moderate hierarchy 
could likely be realized in the neutrino sector. 
For the SeeSaw case, the performances of $H$ and $PA_{\mu \tau}$ are almost equivalent, 
although $PA_{\mu \tau}$ is slightly better.

%
%

\section{Dirac CP phase}

It is also interesting to study the distribution of the Dirac CP-violating phase $\delta$ in the models considered.
At present, there is just a very mild $1\sigma$ preference for $\delta \sim \pi$ in the NH case \cite{Fogli:2012ua}. 

\begin{figure}[t!]
\begin{center}
\begin{tabular}{c}
\includegraphics[width=6.5cm]{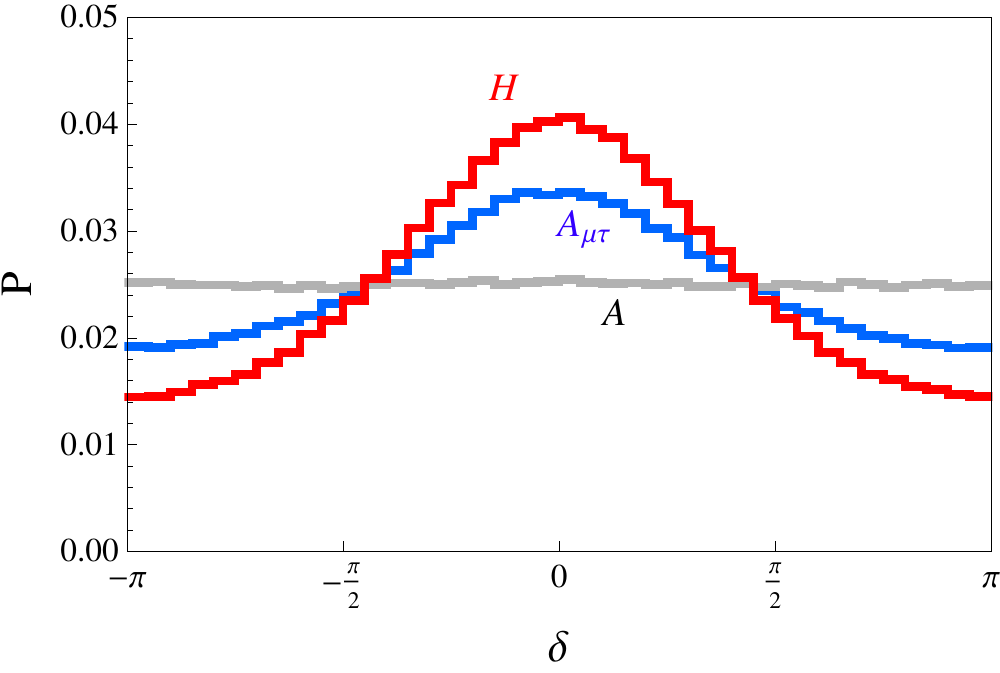}     \,\,\,\,\,\, \includegraphics[width=6.5cm]{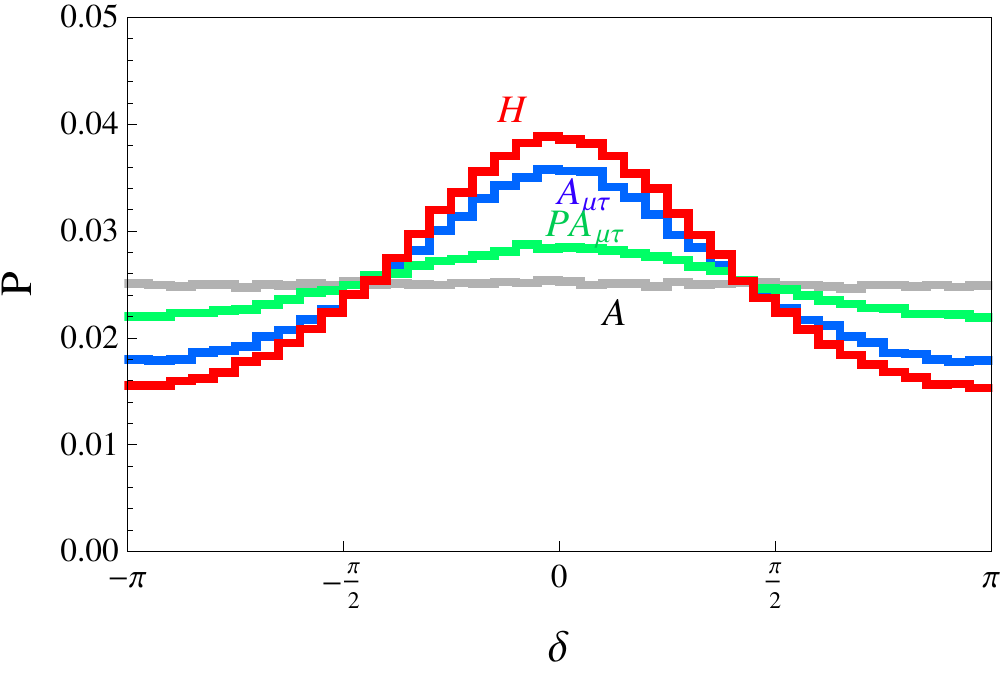}  
\end{tabular}
\end{center}\vspace*{-0.4cm} 
\caption{Probability distributions of $\delta$ without (left column) and with (right column) SeeSaw. 
The modulus (argument) of the complex random coefficients has been generated in the interval $[0.5,2]$ ($[0,2\pi]$) with a flat distribution.
For $A$ and $A_{\mu\tau}$ we considered $\lambda=0.2$, for $H$ and $PA_{\mu \tau}$ we considered $\lambda=0.4$. Note that we considered $\delta \in [-\pi, \pi] $.
}
\label{fig-Dd}
\end{figure}

In order to extract the phase $\delta$, we consider the following combination
\beq
\mathcal{I}=e^{i {\rm Arg}( U_{e3} \, U_{e2}^* \, U_{\mu 3}^* \, U_{\mu 2})},
\eeq
that is an invariant under phase transformation of the fields. This is only one of the possible invariants that can be considered (notice that the imaginary part of $U_{e3} \, U_{e2}^* \, U_{\mu 3}^* \, U_{\mu 2}$ corresponds to the Jarlskog invariant \cite{Jarlskog:1985ht}). Adopting the usual PDG parameterisation of the PMNS matrix, we get 
\begin{equation}
\mathcal{I}\,\,  |U_{\mu 2}| 
= \cos \theta_{23} \, \cos \theta_{12}\, e^{-i\delta} - \, \sin \theta_{23} \,  \sin \theta_{13} \, \sin \theta_{12} \,\,.
\end{equation}
The distributions of $\delta$ are shown in Fig.~\ref{fig-Dd}. As one expects, for the Anarchy model the distribution of $\delta$ is completely flat. On the other hand, for $H$ and $A_{\mu\tau}$ there is a mild preference for a vanishing value of $\delta$, while for $PA_{\mu \tau}$ this preference is even weaker.

%
%

\section{Conclusion}

Over the years there has been a continuous progress in the measurement of neutrino mixing angles culminating recently with the determination of a relatively large value of $\theta_{13}$ and with the indication that $\theta_{23}$ is not maximal (some hints that $\cos\delta \lesssim 0$, with  $\delta$ being the Dirac CP-violating phase in neutrino oscillations, are also emerging). In spite of this remarkable experimental progress the data can still be reproduced by a wide range of theoretical models. At one extreme we have models where the assumed dynamics is minimal and the dominant ingredient is pure chance (Anarchy and its variants) and, at the other extreme, models with a high level of underlying symmetry, like, for example, those based on discrete non-Abelian symmetries (which start at LO with TB or BM mixing). The large value of $\theta_{13}$ and the departure of $\theta_{23}$ from maximal both go in the direction of Anarchy and move away from the TB or BM limits, where $\theta_{13}=0$ and $\theta_{23}$ is maximal. In this note we have made a reappraisal of Anarchy, given the new experimental results. To make connection with quark masses and mixing we have adopted the (SUSY) $SU(5)\otimes U(1)_{\rm FN}$ GUT framework. The Anarchy prototype model has only tenplet charge differences (among the 3 generations) that are non-vanishing, while all pentaplet and singlet charge differences are taken as vanishing. 
Here we argue on the most recent data that the Anarchy ansatz, in the context of $SU(5)\otimes U(1)_{\rm FN}$ models, is simple, elegant and viable but does not provide 
a unique interpretation of the data in that context.
In fact, suitable differences of $U(1)_{FN}$ charges, if also introduced within pentaplets and singlets, lead to distributions that are in much better agreement with the data with the same number of random parameters as for Anarchy. The hierarchy of quark masses and mixing and of charged lepton masses in all cases impose a hierarchy defining parameter of the order of $\lambda_C=\sin{\theta_C}$, with $\theta_C$ being the Cabibbo angle. The weak points of Anarchy ($A$) are that with this ansatz all mixing angles should be of the same order, so that the relative smallness of $\theta_{13}\sim \cO(\lambda_C)$ is not automatic. Similarly the smallness of $r=\Delta m^2_{solar}/\Delta m^2_{atm}$ is not easily reproduced: with no SeeSaw $r$ is of $\cO(1)$, while in the SeeSaw version of Anarchy the problem is only partially alleviated by the spreading of the neutrino mass distributions that follows from the product of three matrix factors in the SeeSaw formula. An advantage is already obtained if Anarchy is only restricted to the 23 sector of leptons as in the $A_{\mu \tau}$ model. In this case, with or without SeeSaw,  $\theta_{13}$ is naturally suppressed and, with a single fine tuning one gets both $\theta_{12}$ large and $r$ small. Actually we have shown that, in the no SeeSaw case, a very good performance is observed in the new $H$ model, where the Anarchy is also abandoned in the 23 sector. In the $H$ model, by taking a relatively large order parameter, one can reproduce the correct size for all mixing angles and mass ratios. In the SeeSaw case, we have shown that the freedom of adopting RH neutrino charges of both signs, as in the $PA_{\mu \tau}$ model, can be used to obtain a completely natural model where all small quantities are suppressed by the appropriate power of $\lambda$. In this model a lopsided Dirac mass matrix is combined with a generic Majorana matrix to produce a neutrino mass matrix where the 23 subdeterminant is suppressed and thus $r$ is naturally small with unsuppressed $\theta_{23}$. In addition also $\theta_{12}$ is large, while $\theta_{13}$ is suppressed. We stress again that the number of random parameters is the same in all these models: one coefficient of $\cO(1)$ for every matrix element. Moreover, with an appropriate choice of charges, it is not only possible to reproduce the charged fermion hierarchies and the quark mixing, but also the order of magnitude of all small observed parameters can be naturally guaranteed. 
Thus finally we agree that models based on chance are still perfectly viable, but we consider Anarchy as a simplest possibility that has to be validated on the data 
in comparison with other similar models and we argue in favor of less chaos than assumed in Anarchy.

%
%
\section*{Acknowledgements}

We recognize that this work has been partly supported by the Italian Ministero dell'Uni\-ver\-si\-t\`a e della Ricerca Scientifica, under the COFIN program (PRIN 2008), by the European Commission, under the networks ``Heptools'', ``Quest for Unification'', ``LHCPHENONET'' and European Union FP7  ITN INVISIBLES (Marie Curie Actions, PITN- GA-2011- 289442) and contracts MRTN-CT-2006-035505 and  PITN-GA-2009-237920 (UNILHC), and by the Te\-ch\-ni\-sche Universit\"at M\"unchen -- Institute for Advanced Study, funded by the German Excellence Initiative. We thank the Galileo Galilei Institute for Theoretical Physics for the hospitality and the INFN for partial support during the completion of this work.

%
%


\providecommand{\href}[2]{#2}\begingroup\raggedright\endgroup

\end{document}